\documentclass[12pt,a4paper,twoside]{article}

\newif\ifpdf
  \ifx\pdfoutput\undefined
  \pdffalse
\else
  \pdfoutput=1
  \pdftrue
\fi

\usepackage[latin1]{inputenc}

\usepackage[english]{babel}

\usepackage{multicol}

\ifpdf

\usepackage{ae}

\usepackage[pdftex]{graphicx}

\usepackage[pdftex]{geometry}
\else

\usepackage[T1]{fontenc}

\usepackage[dvips]{graphicx}

\usepackage[dvips]{geometry}
\fi

\usepackage{dcolumn}
\usepackage{bm}
\usepackage{epsfig}
\usepackage{amsfonts}
\usepackage{amssymb,amscd}

\def\lsim{\raise0.3ex\hbox{$<$\kern-0.75em\raise-1.1ex\hbox{$\sim$}}}
\def\gsim{\raise0.3ex\hbox{$>$\kern-0.75em\raise-1.1ex\hbox{$\sim$}}}

\def\gappeq{\mathrel{\rlap {\raise.5ex\hbox{$>$}}{\lower.5ex\hbox{$\sim$}}}}
\def\lappeq{\mathrel{\rlap{\raise.5ex\hbox{$<$}}{\lower.5ex\hbox{$\sim$}}}}
\def\Toprel#1\over#2{\mathrel{\mathop{#2}\limits^{#1}}}

\def\beq{\begin{equation}}
\def\eeq{\end{equation}}
\def\bea{\begin{eqnarray}}
\def\eea{\end{eqnarray}}

\title{Double diffractive  meson production and the BFKL Pomeron at  $e^+e^-$ colliders}

\author{V.~P. Gon\c{c}alves \thanks{barros@ufpel.edu.br} \\
High and Medium Energy Group (GAME) \\
Instituto de F\'{\i}sica e Matem\'atica,\\ Universidade Federal de Pelotas\\
Caixa Postal 354, CEP 96010-900, Pelotas, RS, Brazil \\
\and
W.~K. Sauter \thanks{sauter@if.ufrgs.br} \\ GFPAE, IF-UFRGS \\
Caixa Postal 15051, CEP 91501-970, \\ Porto Alegre, RS, Brazil}

\date{\today}

\begin{document}
\maketitle

\begin{abstract}

In this Letter we study the double diffractive vector meson production in $e^+e^-$ collisions  assuming the dominance of the BFKL pomeron exchange.  We consider the non-forward solution of the BFKL equation  at high energy and large momentum transfer and  estimate the total cross section for the process $e^+e^- \rightarrow e^+e^- V_1 V_2$ with antitagged $e^+$ and $e^-$, where $V_1$ and $V_2$ can be any two vector mesons  ($V_i = \rho, \omega, \phi, J/\Psi, \Upsilon$). 
The event rates for the future linear colliders  are given. 

\end{abstract}


In the last years several studies has demonstrated that the $e^+ e^-$ colliders offer an excellent opportunity to test the QCD dynamics at high energies (For reviews see e.g. Refs. \cite{kw,reviewhighenergy}).
The simplicity of the initial state and the possibility of study of many different combinations of final states making this process very useful for studying the QCD dynamics in  the limit of high center-of-mass energy $\sqrt{s}$  and fixed momentum transfer $t$. This is the domain where we could expect BFKL Pomeron theory  \cite{BFKL} to be applicable, provided that a hard scale exist which allows to use perturbation theory. 
It has motivated the theoretical and experimental analysis of several reactions. In particular, the  $\gamma^* \gamma^*$ total cross section, using the forward solution of the BFKL equation, was calculated in Refs. \cite{gamagama,gambrod,gamboone}, where the photon virtuality provides  the hard scale. One have that while the two-gluon approximation does not describe the OPAL and L3 data points, the leading order (LO) BFKL prediction lies above the data. First attempts to include the next-to-leading order (NLO) corrections to the BFKL equation are encouraging \cite{kimbrod}. Furthermore, the heavy quark production in $\gamma \gamma$ collisions was estimated   in Refs. \cite{mottim,hansson,per2}, demonstrating that the experimental analysis of  this process can be useful to constrain the QCD dynamics. In this case, the  hard scale is the  photon virtuality and/or the heavy quark mass. A comparison with the data from the L3 collaboration is presented e.g.  in Ref. \cite{per2}.   Another possibility to study the BFKL theory  is the double vector meson production in $\gamma \gamma$ collisions \cite{ginzburg,motyka,motyka_ziaja,per1,d2vm}. In this process and in the general case there can be three large momentum scales - the photon virtuality, $Q^2$, the vector meson mass $M_V$ and the momentum transfer $t$. Recently, the double light vector meson production has been extensively analyzed considering distinct approximations \cite{double,born,enberg,papa}. In particular, the exclusive diffractive process $ \gamma^* \gamma^* \rightarrow \rho \rho$ was calculated at Born level in Ref. \cite{born}, considering the BFKL resummation effects in Ref. \cite{enberg} and the  NLO corrections to the impact factors and BFKL kernel in Ref. \cite{papa}. In these studies the hard scale was provided by the photon virtuality and only the forward solution from the BFKL equation was calculated. On the other hand,  in Ref. \cite{double} we have estimated, for the first time, the double vector meson production at photon colliders considering the non-forward solution of the LO BFKL equation  at high energy and large momentum transfer. It has allowed  to estimate the total and differential cross section for the production of heavy and light vector mesons in real photon interactions. 
   In that study we have restricted our analysis for photon colliders.  In order to estimate the feasibility of this process it is important also compute the corresponding $e^+ e^-$ cross sections. In this letter we extend our previous analyzes  and calculate the double diffractive vector meson production in $e^+e^-$ collisions  assuming the dominance of the BFKL pomeron exchange and estimating the event rates for the future linear colliders.



\begin{figure}[t]
\centerline{\psfig{file=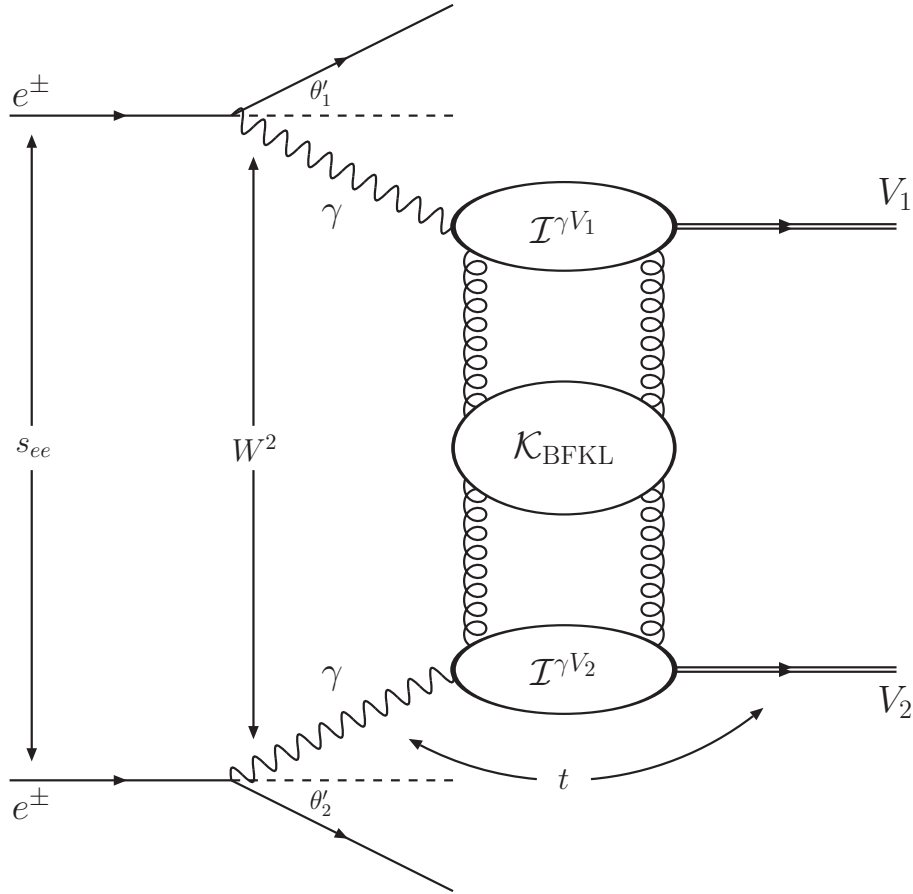,width=120mm}}
 \caption{Impact factor representation for the $e^+e^- \rightarrow e^+e^-V_1V_2$ process.}
\label{fig1}
\end{figure}

Two-photon processes at future high energy linear colliders can be measured either as in a storage ring, via photon emission from the lepton beams, according to a Weizs\"acker-Williams (WW) energy distribution, or using a linear collider in photon collider mode (For a recent review see, e.g. Refs. \cite{nisius,serbo}). In the latter case the high energy electron beam is converted into a high energy photon beam, by backscattering of photons off an intense laser beam, just before the interaction point. Considering $e^+e^-$ collisions, current conservation and the small photon virtuality lead to a factorization of the lepton scattering cross sections into the photon spectrum in the lepton and the hard photon scattering cross section \cite{review}. 
For the $e^+e^- \rightarrow e^+e^- V_1 V_2$ process, the cross section will be given by (see Fig. \ref{fig1})
\bea
\sigma_{e^+e^- \rightarrow e^+e^- V_1 V_2} (\sqrt{s_{ee}})= \int dx_a dx_b dt f_{\gamma/e}(x_a) f_{\gamma/e}(x_b) \frac{d \sigma_{\gamma \gamma \rightarrow V_1 V_2}}{dt} (\hat{s}) \,\,,
\label{sigtee}
\eea
where $s_{ee}$ is the squared $e^+ e^-$ center of mass energy, $x_a$ and $x_b$ denote the fractions of longitudinal momentum of the lepton $a$ and $b$ that are carried by the corresponding photons and $\hat{s}=x_a x_b s_{ee}$ is the center of mass energy of the $\gamma \gamma \rightarrow V_1 V_2$ subprocess. 
In the  Weizs\"acker-Williams (WW) approximation \cite{ww}, the energy spectrum of the exchanged photons is given by 
\bea
f_{\gamma/e}(x) = 
\frac{\alpha_{\rm em}}{2\pi}\left\{ \frac{1+(1-x)^2}{x} \ln\left(\frac{Q^2_{max}}{Q^2_{min}}\right) - 2m^2 x \left[\frac{1}{Q^2_{min}} - \frac{1}{Q^2_{max}} \right] \right\}.
\eea
The subleading nonleading logarithmic terms modify the cross section typically by $5\%$ \cite{frixione}. $\alpha_{\rm em}$ is the electromagnetic coupling constant, $x = E_{\gamma}/E$ is the energy fraction transferred from the electron to the photon, $m$ is the electron mass. Moreover, $Q^2_{min} = (m^2 x^2)/(1-x)$ and $Q^2_{max} = E^2 (1-x)  \theta^2_{max}$ is the maximal photon virtuality for electron scattering angles below $\theta_{max}$. This angle can be determined by tagging the outgoing electron in the forward direction or by requiring that it be lost in the beam pipe (antitagging). Following Ref. ~\cite{motyka} we  have that the Eq. (\ref{sigtee}) can be rewritten as
\bea
\sigma_{e^+e^- \rightarrow e^+e^- V_1 V_2} (\sqrt{s_{ee}})= \int^{1}_{0} dx_a \int^{1}_{0} dx_b\ \Theta(\hat{s} - \hat{s}_{min}) f_{\gamma/e}(x_a) f_{\gamma/e}(x_b) \sigma_{\gamma \gamma \rightarrow V_1 V_2}(\hat{s}),
\label{sigtee2}
\eea
where $\Theta$ is the step function, $\hat{s}_{min}$ is the threshold on  hadron production which assures that the $\gamma \gamma$ system is in the high energy region where the BFKL dynamics is expected to be valid (see below), and
\bea
\sigma_{\gamma \gamma \rightarrow V_1 V_2}(\hat{s}) = \int_{t_{min}}^{\infty}\ dt \,\frac{d \sigma_{\gamma \gamma \rightarrow V_1 V_2}}{dt} (\hat{s})
\label{sigtot}
\eea
is the total cross section of the photon-photon subprocess. Moreover, $t_{min}$ is a cut-off in the momentum transfer (see discussion below).

At high energies the double diffractive meson production is expected to be described in terms of the BFKL Pomeron \cite{ginzburg,motyka,motyka_ziaja,per1,d2vm,double,born,enberg,papa}. At  leading logarithmic approximation, it  corresponds to a sum of ladder diagrams with reggeized gluons along the chain, with the sum being  described by the BFKL equation  \cite{BFKL}.  Considering the impact factor representation and the BFKL dynamics, the differential cross section for the $\gamma \gamma \rightarrow V_1 V_2$ will be  expressed  by (For details see Ref. \cite{double})
\begin{equation}
\frac{d\sigma (\gamma \gamma \rightarrow V_1 V_2)}{dt} \; = \;
\frac{16\pi}{81 t^4}
|{\mathcal{F}}_{\mathrm{BFKL}}(z,\tau)|^{2}\,\,,
\label{dsdtgq}
\end{equation}
where the BFKL amplitude ${\mathcal{F}}_{\mathrm{BFKL}}$, in the leading logarithm approximation (LLA) and lowest conformal spin ($n=0$), is given by~\cite{Lipatov}
\begin{equation}
\label{BFKLa}
{\mathcal{F}}_{\mathrm{BFKL}}(z,\tau)=\frac{t^{2}}{(2 \pi)^{3}}\int d\nu \frac{\nu ^{2}}{(\nu ^{2}+1/4)^{2}}e^{\chi (\nu )z}I_{\nu }^{\gamma V_1}(Q_{\perp })I^{\gamma V_2}_{\nu }(Q_{\perp })^{\ast }.
\end{equation}
In the Eqs. (\ref{dsdtgq}) and (\ref{BFKLa}), $z = [3\alpha_{s}\ln ( \hat{s}/\Lambda^{2})]/(2\pi) $, $\tau = |t|/(M_{V}^{2}+ Q_{\gamma}^{2})$, $M_{V}$ is the mass of the vector meson, $Q_\gamma$ is the photon virtuality, $\Lambda^{2}$ is a characteristic  scale related to $M_V^2$  and $|t|$, and  $Q_{\perp}$ is the momentum transfered, $t=-Q_{\perp}^2$, (the subscript denotes two dimensional transverse vectors). Furthermore,   
\begin{equation}
\chi (\nu )=4{\mathcal{R}}\mathrm{e}\biggl (\psi (1)-\psi \bigg (\frac{1}{2}+i\nu \bigg )\biggr )
\end{equation}
is proportional to the BKFL kernel eigenvalues~(See e.g. Ref. \cite{Jeff-book}), with $\psi(x)$ being the digamma function. 
 The quantities $I_{\nu }^{\gamma V_i}$ are given in terms of the impact factors ${\mathcal{I}}_{\gamma V_i}$ and the BFKL eigenfunctions as follows \cite{FR},
\begin{eqnarray}
I_{\nu}^{\gamma V_i}(Q_{\perp }) & = & -{\mathcal{C}_i}\, \alpha_s \frac{16\pi}{Q_{\perp }^{3}}\frac{\Gamma (1/2-i\nu )}{\Gamma (1/2+i\nu )}\biggl (\frac{Q_{\perp }^{2}}{4}\biggr )^{i\nu }\int _{1/2-i\infty }^{1/2+i\infty }\frac{du}{2\pi i}\biggl (\frac{Q_{\perp }^2}{4 M_{V_i}^2}\biggr )^{1/2+u}\\
 &  & \times\frac{\Gamma ^{2}(1/2+u)\Gamma (1/2-u/2-i\nu/2)\Gamma (1/2-u/2+i\nu/2)}{\Gamma (1/2+u/2-i\nu /2)\Gamma (1/2+u/2+i\nu /2)},\nonumber \label{IV} 
\end{eqnarray}
where the constant ${\mathcal{C}_i}$ may be related to the vector meson leptonic decay width, $\mathcal{C}^{2}_i\;=\;3\Gamma_{ee}^{V_i}M_{V_i}^{3}/\alpha_{\mathrm{em}}$.

The differential cross section can be directly calculated substituting the above expression in Eq. (\ref{BFKLa}) and evaluating numerically the integrals.
Following  Ref. \cite{double}, we will assume the non-relativistic approximation of the meson wave functions, $\alpha_s = 0.2$  and $\Lambda^2 = \beta_1 M_{V_1}^2 + \beta_2 M_{V_2}^2 + \gamma |t| $, with  $\beta_1 = \beta_2 = 1/2$ and  $\gamma = 0$.  
As the differential cross section is proportional to $\alpha_s^4$ and the energy dependence is determined by the variable $z$, which is dependent of the $\alpha_s$ and $\Lambda$ values, we have a strong dependence of our predictions on the choice of these parameters (For a more detailed discussion see Ref. \cite{double}). A similar dependence is expected for the predictions presented in this letter. These uncertainties will be reduced considering in the calculations  the NLO corrections to the impact factors and non-forward BFKL kernel.

In what  follows we calculate the total cross section for the process $e^+e^- \rightarrow e^+e^-V_1V_2$ with  $V_i={\rho,\,\omega,\,\phi,\,J/\psi,\,\Upsilon}$ and different values of the center of mass energy ($\sqrt{s_{ee}} =$  200, 500, 1000 e 3000 GeV). Moreover, we will assume $\theta_{max} = $ 30 mrad. 
As our calculations for the $\gamma \gamma \rightarrow V_1 V_2$ subprocess are only valid at high energies we should to exclude the region where the BFKL dynamics  is expected to break down. This restriction is present in Eq. (\ref{sigtee2}) in terms of the theta function, where we have defined 
 a minimum value for the center of mass energy of the hard subprocess, $\hat{s}$. Here we choose $\sqrt{\hat{s}_{min}} = W_{min} = 20$ GeV, independently of the hadronic final state produced. Furthermore, for double light meson production, the hard scale for the $\gamma \gamma \rightarrow V_1 V_2$ subprocess is provided by the momentum transfer. Consequently, in the calculation of the total cross section [Eq. (\ref{sigtot})] 
 a lower cut-off in the $t$-integration is necessary in order to minimize non-perturbative (Soft Pomeron) contributions. We choose in this case $|t|_{min}=1$ GeV$^2$, which is reasonable considering that the HERA data for the photoproduction of light vector mesons in this kinematic region are  quite well described using a similar approach \cite{FP,jhep}. On the other hand, we assume $|t|_{min}=0$ when at least a heavy meson is produced, since in this case we have a  hard scale present which justifies the perturbative calculations in the low-$t$ ($t \approx 0$) region. It is important to emphasize that a Soft Pomeron contribution is not included in our calculations (See e.g. Ref. \cite{ddr}).

Our predictions for the double diffractive meson production in  $e^+e^-$ collisions are presented in the Tables \ref{tabfklh} and \ref{tabfkll} considering the non-forward solution of the BFKL equation. For comparison we also show the results obtained at the Born level, which corresponds to the two gluon exchange mechanism. It is important to emphasize that although the Born term implies an energy independent cross section at the photon level, its predictions for the $e^+ e^-$ cross sections have a weak energy dependence due to the photon flux factor. As already observed in Ref. \cite{double}, the cross sections decrease when a heavier vector meson is considered. Moreover, the BFKL dynamics implies an enhancement of the cross section in comparison to the two-gluon exchange predictions, which can reach two orders of magnitude, depending of the energy and mesons involved in the reaction. 
Previous estimates for the double $J/\Psi$ and $\rho J/\Psi$ production in $e^+e^-$ collisions has been obtained in Refs. \cite{motyka}  and  \cite{motyka_ziaja}, respectively. The production of other combinations of vector mesons in the final state considering the BFKL dynamics  are estimate here for the first time.   In Ref. \cite{motyka} the authors have solved the BFKL equation taking into account dominant non-leading effects which come from the requirement that the virtuality of the exchanged gluons along the gluon ladder is controlled by their transverse momentum squared and have estimated the double $J/\Psi$ production at LEP2 energy.   We have that our predictions are similar with those obtained there if we consider the same experimental cuts and energy. A more detailed comparison of the energy dependence is not possible, since it is not analyzed in Ref. \cite{motyka}. On the other hand,  the $\rho J/\Psi$ production in $e^+ e^-$ collisions was previously estimated  in Ref. \cite{motyka_ziaja}, where this  process  was proposed as a probe of the gluon distribution on the meson $xG^{\rho}$ and, by consequence, a constrain for the photon structure.
Our predictions agree with the results presented there, only presenting a steeper energy growth. It is expected since we are using the LO solution of the BFKL equation, while in Ref. \cite{motyka_ziaja} the energy dependence is determined by the gluon distribution of the light meson which has a smaller intercept.  

\begin{table}[t]
\begin{center}
{ \footnotesize
\begin{tabular}{||c| r@{.}l@{~}r@{.}l| r@{.}l@{~}r@{.}l| r@{.}l@{~}r@{.}l| r@{.}l@{~}r@{.}l||}
\hline
\hline
  &  \multicolumn{4}{c|}{$\sqrt{s_{ee}} = 200$ GeV} & \multicolumn{4}{c|}{$\sqrt{s_{ee}} = 500$ GeV} & \multicolumn{4}{c|}{$\sqrt{s_{ee}} = 1000$ GeV}   & \multicolumn{4}{c||}{$\sqrt{s_{ee}} = 3000$ GeV}   \\
\hline
$\rho J/\Psi$ & 0&90 &(0&015)  & 5&80 &(0&049) & 21&87 &(0&097) & 178&19 &(0&22)  \\
\hline
$\phi J/\Psi$ & 0&11 &(0&0023) & 0&69 &(0&0073) & 2&58 &(0&014) & 20&77 &(0&033) \\
\hline
$\omega J/\Psi$ & 0&075 &(0&0013) & 0&48 &(0&0041) & 1&85 &(0&0081)  & 15&03 &(0&019) \\
\hline
$J/\Psi J/\Psi$ & 0&045 &(0&0021) & 0&27 &(0&0066) & 0&98 &(0&012)  & 7&56 &(0&031) \\
\hline
$\rho \Upsilon$ & 0&0013 &(0&000055) & 0&0093 &(0&00017) & 0&036 &(0&00034) & 0&31 &(0&00080) \\
\hline
$\omega \Upsilon$ & 0&00011 &(0&0000055) & 0&00078 &(0&000017) & 0&0030 &(0&000034) & 0&026 &(0&000080) \\
\hline
$\phi \Upsilon$ & 0&0002 &(0&000011) & 0&0013 &(0&000034) & 0&0050 &(0&000068) & 0&040 &(0&00016) \\
\hline
$J/\Psi \Upsilon$ & 0&00025 &(0&000027) & 0&0015 &(0&000086) & 0&0052 &(0&00017) & 0&038 &(0&00040) \\
\hline
$\Upsilon \Upsilon$ & 0&0000072 &(0&0000014) & 0&000038 &(0&0000045) & 0&00012 &(0&0000088) & 0&0008 &(0&000020) \\
\hline
\hline
\end{tabular} }
\end{center}
\caption{The double vector meson production cross sections in
$e^+ e^-$ processes at different energies,  $|t|_{min} = 0$ and $\theta_{max} = 30 $ mrad, assuming the  BFKL Pomeron (Two-gluon) exchange. Cross sections are given in pb.  } 
\label{tabfklh}
\end{table}


\begin{table}[t]
\begin{center}
\begin{tabular}{||c| r@{.}l@{~}r@{.}l| r@{.}l@{~}r@{.}l| r@{.}l@{~}r@{.}l| r@{.}l@{~}r@{.}l||}
\hline
\hline
  &  \multicolumn{4}{c|}{$\sqrt{s_{ee}} = 200$ GeV} & \multicolumn{4}{c|}{$\sqrt{s_{ee}} = 500$ GeV} & \multicolumn{4}{c|}{$\sqrt{s_{ee}} = 1000$ GeV}   & \multicolumn{4}{c||}{$\sqrt{s_{ee}} = 3000$ GeV}   \\
\hline
$\rho \rho$ & 0&18 &(0&035)  & 1&03 &(0&11) & 3&60 &(0&21) & 26&62 &(0&51)  \\
\hline
$\rho \phi$ & 0&033 &(0&0053) & 0&19 &(0&016) & 0&66 &(0&032) & 4&87 &(0&077) \\
\hline
$\rho \omega$ & 0&015 &(0&0030) & 0&088 &(0&0093) & 0&309 &(0&018)  & 2&28 &(0&043) \\
\hline
$\phi \phi$ & 0&0067 &(0&00084) & 0&038 &(0&0026) & 0&130 &(0&0051) & 0&951 &(0&012) \\
\hline
$\phi \omega$ & 0&0029 &(0&00044) & 0&016 &(0&0013) & 0&057 &(0&0027) & 0&41 &(0&0064) \\
\hline
$\omega \omega$ & 0&0013 &(0&00025) & 0&007 &(0&00080) & 0&026 &(0&0025)  & 0&19 &(0&0036) \\
\hline
\hline
\end{tabular}
\end{center}
\caption{The double light vector meson production cross sections in
$e^+ e^-$ processes at different energies, $|t|_{min} = 1$ GeV$^2$  and $\theta_{max} = 30$ mrad, assuming the  BFKL Pomeron (Two-gluon)  exchange. Cross sections are given in pb.  } 
\label{tabfkll}
\end{table}

\begin{table}[t]
\begin{center}
\begin{tabular}{||c|r@{.}l|r@{.}l|r@{.}l||}
\hline
\hline
  & \multicolumn{2}{c|}{TESLA} & \multicolumn{2}{c|}{CLIC} & \multicolumn{2}{c||}{ILC}  \\
\hline
$\rho \rho$  & 350000&0 & 206000&0 & 220000&0  \\
\hline
$\rho J/\Psi$  & 1970000&0 & 1160000&0 & 1270000&0   \\
\hline
$J/\Psi J/\Psi$  & 92000&0 & 54000&0 & 59000&0  \\
\hline
$\Upsilon \Upsilon$ & 13&0 & 8&0 & 9&0 \\
\hline
\hline
\end{tabular}
\end{center}
\caption{Number of events per year for double vector meson production at  TESLA, CLIC and ILC expected luminosities ($\sqrt{s_{ee}}$ = 500 GeV and $\theta_{max} = 30$ mrad).} 
\label{numev}
\end{table}

Finally, lets estimate the expected number of events of some of the processes calculated in this paper for the future linear colliders \cite{tesla,snow,clic}.  For $e^+ e^-$ collisions with center of mass energies equal to $\sqrt{s_{ee}} = 500$ GeV, luminosities of order ${\cal{L}} = 340$, 200 and 220 $fb^{-1}/year$ are expected at TESLA, CLIC and ILC, respectively. 
 In Table \ref{numev} we present our predictions for the number of events per year for the $\rho \rho$,  $\rho J/\Psi$,  $J/\Psi J/\Psi$ and $\Upsilon \Upsilon$ production. 
For double $\rho$ production our estimate is conservative, since we consider a cut in the momentum transfer $t$ ($|t|_{min} = 1$  GeV$^2$) and    the Soft Pomeron contribution is not included in our calculations. As expected, the number of events is smaller than those obtained for a  $\gamma \gamma$ collider \cite{double}. However, we still predict a large number of events related to double meson production in $e^+ e^-$ collisions, allowing future experimental analyses, even if the acceptance for  vector meson detection were low. Consequently, we believe that this process could be used to constrain the QCD dynamics at high energies.

As a summary, we have studied the  double diffractive vector meson production in $e^+e^-$ collisions  assuming the dominance of the BFKL Pomeron exchange  and estimated the total cross sections for the future linear colliders. Our results indicate that it may be possible to perform a sucessful experimental analysis of this process. It is an important results since, as already pointed out in Ref. \cite{enberg}, this exclusive diffractive reaction may become the best tool to investigate the perturbative picture of the BFKL Pomeron. In particular, we expect that in this process the impact of the NLO corrections can be evaluated and the various approaches proposed in the literature can be constrained.

\section*{Acknowledgments}
  W. K. S. thanks the support of the High Energy Physics Phenomenology Group at the Institute of Physics, GFPAE IF-UFRGS, Porto Alegre and the hospitality of IFM-UFPel  where this work was accomplished. This work was partially financed by the Brazilian funding agencies CNPq and FAPERGS.

\end{document}